\documentclass[english,twoside,a4paper,12pt]{article}

\usepackage[english]{babel}
\usepackage{epsfig}
\usepackage{fancyhdr} 
\usepackage{amsmath,amssymb}
\usepackage{amscd} 
\usepackage{amsmath}
\usepackage[utf8]{inputenc}
\usepackage[T1]{fontenc}
\usepackage{xcolor}
\usepackage{cite}
\usepackage{amsmath}
\usepackage{amsfonts}
\usepackage{graphicx}
\usepackage{subfigure}
\usepackage{a4wide}
\usepackage{amssymb}
\usepackage{fancyhdr}
\usepackage{mathrsfs}
\usepackage[toc,page]{appendix}

\begin{document}

\title{A special class of solutions in $F(R)$-gravity}

\author{ 
Marco Calzà$^1$\footnote{E-mail address: marco.calza89@gmail.com},\,\,\,
Massimiliano Rinaldi$^{1,2}$\footnote{E-mail address: massimiliano.rinaldi@unitn.it},\,\,\,
Lorenzo Sebastiani$^{1,2}$\footnote{E-mail address: lorenzo.sebastiani@unitn.it}\\
\\
\begin{small}
$^1$Dipartimento di Fisica, Universit\`a di Trento,Via Sommarive 14, 38123 Povo (TN), Italy
\end{small}\\
\begin{small}
$^2$TIFPA - INFN,  Via Sommarive 14, 38123 Povo (TN), Italy
\end{small}
}

\date{}

\maketitle

\abstract{
\noindent We consider a special class of vacuum $F(R)$-modified gravity models. The form of their  
Lagrangian is such that  the field equations are trivially satisfied when the Ricci scalar is constant. There are many interesting $F(R)$-models 
for inflation and dark energy that fall in this class. However, little is known outside the domain of cosmology therefore 
we aim to explore the class of solutions that are static and spherically symmetric. After some general considerations, we investigate 
in more detail black hole solutions, traversable wormhole metrics and, finally, configurations that can match the anomalous  rotation curves of galaxies. 
}

\section{Introduction}

Despite
the successful results obtained by General Relativity (GR) in describing the
Universe and the Solar System, many open problems remain and
it is well accepted the idea that Einstein's gravity may be
not the ultimate theory of gravity. 
The origin of the cosmic acceleration and the nature
of the ``dark'' contents of the Universe is still unknown
and can be the effects of some modifications of the Einsteinian theory
of gravity itself. In modified gravity the Hilbert-Einstein action of GR is replaced by a more general 
function of the curvature invariants. The simplest models are given by the so-called  $F(R)$ gravity, 
where the gravitational Lagrangian is a function of the 
Ricci scalar $R$ instead of just $R$ (see Refs.~\cite{Od1,Od2, Capo1, Capo2,Od3,Od4,Od5,Od6,Od7} and references therein for some reviews).
With these modifications of the gravitational Lagrangian, some observed features of our Universe, like the early-time inflation or the late-time accelerated expansion, become
an effect of the curvature, without the introduction of any dark fluids or special scalar fields. Finally, specific deformations of the usual Einstein-Hilbert Lagrangian are motivated by quantum gravitational corrections, which might leave an observational imprint during the inflationary era \cite{modgravinf1,modgravinf2,modgravinf3}.

In the context of cosmology, $F(R)$ gravity has raised a lot of interest as it represents a promising alternative to scalar inflation and/or dark energy. This said, if we take for good $F(R)$ gravity, we must ensure that it works also on much smaller scales and energies.  For example, in order to verify the consistence of a $F(R)$ theory with the standard
Solar System tests, it is important to study the possible deviations from the usual Schwarzschild metric. However, since the equations of motion of $F(R)$ 
are quite involved, the problem of finding spherically symmetric solutions in $F(R)$ gravity is a formidable task that has attracted a flurry research activity  ~\cite{Cap, Clifton, Multa, Dimakis:2017tvb, Nojiri:2017ncd, Cisterna:2015iza}.

In this paper, we consider a special class of $F(R)$-gravity models whose Lagrangians satisfy two constraints, namely $F(R_{0})=0$ and $F_{R}(R_{0})=0$, 
where $F_{R}$ denotes the first derivative of $F(R)$ with respect to $R$ and $R_{0}$ is a constant (eventually vanishing) value for the Ricci scalar. 
As we will see below, if a Lagrangian fulfills these requirements then the vacuum equations of motion are automatically satisfied. 
Note that General Relativity does not belong to this class of theories since its vacuum equations of motion are identically satisfied by any metric with $R=0$ but $F_{R}(R)=1$.

Looking for models that satisfy these conditions is a strategy that allows finding new solutions to the equations of motion of $F(R)$ gravity without solving them explicitly. 
We will see that several models for the early-time inflation or late-time accelerated expansion fall in this class of theories.
If on one side this models have been deeply investigated
at the cosmological level, on the other side there are not many solutions that can describe black holes, wormholes or compact objects in general, 
including the outer regions of spiral galaxies. 
Thus, we focus on the class of static and spherically symmetric solutions with constant (or vanishing) Ricci scalar that also satisfy the general vacuum $F(R)$ equations of motion.

The paper is organized in the following way. In Section {\bf 2} we introduce our class of $F(R)$-theories with exact solutions for constant Ricci scalar curvature and
we discuss several examples of Lagrangians. In Section {\bf 3} we study the black holes solutions. Sections {\bf 4} and Section {\bf 5} are devoted to the analysis 
of solutions for traversable wormholes and for the profile of rotation curves of galaxies. Conclusions are given in Section {\bf 6}.

In this paper, we set $\kappa^2=8\pi G_N$, where $G_{N}$ is the  Newton's constant.

\section{The model Ansatz}

The gravitational action of $F(R)$-gravity in four dimension reads,  
\begin{equation}
 I=\int_\mathcal{M} d^4 x\sqrt{-g}\left[\frac{F(R)}{2\kappa^2}+\mathcal{L}_m\right]\,, \label{action}
\end{equation}
where $\mathcal M$ is a four-dimensional space-time manifold, $g\equiv g(x^\mu)$ is
the determinant of the metric tensor $g_{\mu\nu}(x^\mu)$, $F(R)$ is a function of the Ricci scalar $R$, and $\mathcal{L}_m$ is the matter Lagrangian.
When $F(R)=R$ we recover the Hilbert-Einstein action of General Relativity. The field equations obtained from \eqref{action} are 
\begin{equation}
F_R(R)G_{\mu\nu}=\kappa^2 T_{\mu\nu}+\frac{1}{2}g_{\mu\nu}\left[F(R)-RF_R(R)\right]
+(\nabla_{\mu}\nabla_{\nu}-g_{\mu\nu}\Box)\,F_R(R)\,,
\label{fieldequationF(R)}
\end{equation}
where $F_{R}(R)\equiv dF(R)/dR$. Also,  $T_{\mu\nu}$ is the stress-energy tensor associated to the matter contents of the space-time. 

Some general remarks are in order. First of all, when $R=T_{\mu\nu}=0$ one expects to recover the Ricci-flat class of vacuum solutions of GR. Therefore, for a generic but viable $F(R)$ model the condition $F(0)=0$ is required. In addition, in the presence of standard matter or radiation, 
one must implement the condition $0<F_R(R)$ to avoid gravity becoming repulsive. This is evident from the expression of the effective gravitational coupling that reads $G_{\mathrm{eff}}:=G_N/F_R(R)$.

A crucial point, at the heart of this paper, is the following: when $T_{\mu\nu}=0$  the field equations \eqref{fieldequationF(R)} are automatically satisfied if
 \begin{equation}
F(R_0)=0\,,\quad F_R(R_0)=0\,,\label{Ansatz} 
\end{equation}
where $R=R_0$ is a real positive or negative constant. In the following, we show that looking for models, which fulfill (\ref{Ansatz}) for some values of the curvature $R_0$, is reasonable since they can unveil some cosmological scenarios that cannot be described 
in the frame of standard GR. 

As a concrete example, let us consider the models 
\begin{eqnarray}
F(R)&=& R-2\lambda\left(1-\text{e}^{-{R}/(2\lambda)}\right)\,,\nonumber\\ 
F(R)&=& R-2\lambda\tanh\left(\frac{R}{2\lambda}\right)\,,\label{class0}
\end{eqnarray}
which satisfy the conditions (\ref{Ansatz}) for $R_0=0$. 
If we identify $\lambda$ with the experimental value of the cosmological constant $\Lambda$, we recover some of the cases of exponential modified gravity discussed in
Refs.~\cite{nostriexp, nostriexp2, altriexp} or Refs.~\cite{new3, new1, new2} for recent works (for instance, the exponential model in \cite{new3} satisfies (\ref{Ansatz})
with $R_0=0$ when $A=B$ and $C=4A D$). Exponential gravity belongs to the class of so-called ``one-step models'', which reproduce an 
effective cosmological constant for large values of curvature, while, in the flat limit, they make it vanishing (see e.g.  Refs.~\cite{StaroModel, HS, HSOd, Battye}). 
For instance, for the second of models \eqref{class0}, we see that when $1\ll R/(2\lambda)$, $F(R)\simeq R-2\lambda$
and $F_R(R)\simeq 1$ avoiding anti gravitational effects.
Instead,  $R=0$ implies $F(0)=0$ and we get the Minkowski space-time of Special Relativity.

Note that, because of conditions \eqref{Ansatz}, the above examples admit, as a solution, all metrics such that $R=0$. On the contrary, 
in standard GR only the Schwarzschild solution (or Minkowski) is a vacuum solution of the field equations $G_{\mu\nu}=0$.

The deviations  from the Einstein's theory that take
place  at large curvatures are interesting in the context of the phenomenology of the primordial Universe. 
Within this frame, a general class of models which satisfy (\ref{Ansatz}) is given by
\begin{equation}
F(R)=\gamma(R-R_0)^n=\gamma\sum_{k=0}^{n}\frac{n!}{k!(n-k)!}R^{n-k}(-R_0)^k\,,\label{class1}
\end{equation}
which can be conveniently recast  as 
\begin{equation}
F(R)=\gamma\frac{n!}{(n-1)!}(-R_0)^{n-1}R+\gamma (-R_0)^n+\gamma \sum_{k=0}^{n-2} \frac{n!}{k!(n-k)!}R^{n-k}(-R_0)^k\,.\label{FRexpl}
\end{equation}
Here, $\gamma$ is a dimensional parameter and $2\leq n$. The linear term reproduces the Hilbert-Einstein part of the action provided
\begin{equation}
 (-R_0)=\left[\frac{(n-1)!}{\gamma n!}\right]^{\frac{1}{n-1}}\,.\label{c0}
\end{equation}
If we assume that $\gamma\ll 1$,  the powers of $R$  play a fundamental role only at a  curvature of the order of $R_{0}$. However, a potentially very large cosmological constant $\gamma(-R_0)^n$ emerges in the action so some mechanism to suppress it at small curvature must
be introduced. A simple way to do so is to make the replacement $\gamma\rightarrow g(R)\gamma$, $g(R)$ being a suitable function of the curvature, in front of $(-R_0)^n$.
Thus,
\begin{equation}
F(R)=R+\gamma g(R) (-R_0)^n+\gamma \sum_{k=0}^{n-2} \frac{n!}{k!(n-k)!}R^{n-k}(-R_0)^k\,,  
\end{equation}
where $R_0$ is given by (\ref{c0}). The function $g(R)$ must be chosen so that
\begin{equation}
g(R_0)=1\,,\quad g_R(R_0)=0\,,\quad g(R\ll R_0)\simeq 0\,. 
\end{equation}
The first two conditions are necessary to still fulfill (\ref{Ansatz}), while the third condition leads to GR in the limit of small curvature $F(R\ll R_0)\simeq R$.
We further require that
\begin{equation}
g(0)=0\,,\quad g_R(R\ll R_0)\simeq 0\,, 
\end{equation}
so that $F(0)=0$ (and Minkowski space is recovered for vanishing curvature) and $F_R(R\ll R_0)\simeq 1$ so anti-gravitational effects are avoided when matter and radiation are present. Notable examples which satisfy these requirements are 
\begin{eqnarray} 
g(R)&=&\left(\frac{\text{e}^{-\beta R_0^{2m}}-\text{e}^{-\beta(R-R_0)^{2m}}}{\text{e}^{-\beta R_0^{2m}}-1}\right)\,,\quad 0<\beta\,, 1\leq m\,, \nonumber\\
g(R)&=&\left(\frac{\tanh\beta R_0^{2m}-\tanh\beta (R-R_0)^{2m}}{\tanh\beta R_0^{2m}}\right)\,,\quad 0<\beta\,, 1\leq m\,,
\end{eqnarray}
where $\beta\,,m$ are positive parameters. These functions represent in fact a sort of extension of exponential gravity (useful in the late Universe) towards the inflationary scenario,
since the presence of a large effective cosmological constant at large curvature can drive inflation. When the cosmological constant is suppressed by the mechanism explained above, the Universe gracefully exits inflation to enter the standard radiation epoch governed by Einstein's gravity.
Nevertheless, higher curvature terms are still necessary to support the reheating and protect the theory against finite-future time singularities~\cite{classificationSingularities, OdSing, GBSingularities}. 

For these reasons, classes of models of the form of (\ref{class1}) have interesting applications in the phenomenology of the early Universe. For example, 
the simplest case, with $n=2$,
leads to the appearence of a quadratic correction to Einstein's gravity. Such a term can support a Starobinsky-like inflation~\cite{Staro} when $0<\gamma$ (the emerging of the 
cosmological constant eventually fixes the curvature of the de Sitter (dS) solution, unless one does not want to consider asymptotic solutions with $R_0\ll R$), 
while when $\gamma<0$ the cosmological bounce replaces the singularity of the 
Big Bang~\cite{Odbounce, miobounce}. Moreover, this models automatically solve the field equations for every solution with constant curvature $R=R_0$.

Other examples of Lagrangians that satisfy condition (\ref{Ansatz}) and whose implementation makes sense at large values of curvatures $R\simeq R_0$ 
are
\begin{eqnarray}
F(R)&=&\gamma\left[\cos\left[\beta(R-R_0)\right]-1\right]\,,\nonumber\\ 
F(R)&=&\gamma\left[\sin\left(\frac{\pi R}{2 R_0}\right)-1\right]\,,\nonumber\\
F(R)&=&\gamma\left[\cosh\left[\beta(R-R_0)\right]-1\right]\,,
\end{eqnarray}
where $\gamma\,,\beta$ are dimensional parameters. In these cases, the GR limit is found by Taylor expansions for $|R-R_0|\ll 1$. 

Modified theories of gravity which reproduce the cosmological accelerations typical of inflation and dark energy are certainly interesting and
the models discussed in this section have been thoroughly studied for these cases. 
As a general feature, they admit every class of solutions with constant Ricci scalar $R=R_0$.
At the cosmological level, when one describes the homogeneous and isotropic Universe by using the Friedmann-Robertson-Walker metric,
the most interesting and well-known solutions with zero or constant curvature are the radiation-dominated solution and the de Sitter space-time, respectively. 

In the context of static and spherically symmetric solutions, especially the ones describing black holes (BHs), the models described in this Section have not been investigated so it may be interesting to fill this gap. Therefore, in what follows, we will investigate  static and spherically symmetric metrics with constant (eventually zero) Ricci scalar as exact solutions
of the models discussed above.

\section{Static Spherically Symmetric solutions with constant Ricci scalar and black holes}

The metric of a generic static spherically symmetric (SSS) space-time is characterized by two functions of the radial coordinate $r$ only, say $\alpha=\alpha(r)$ and $B=B(r)$. Its general form is
\begin{equation}
 ds^2=-\text{e}^{2\alpha(r)}B(r)+\frac{dr^2}{B(r)}+r^2d\Omega^2_k\,,\quad d\Omega_k^2= \frac{1}{1-{k \rho^2}} d\rho^2 +\rho^2 d\phi^2
 \,.\label{metric}
\end{equation}
Here, $d\Omega^2_k$ is the metric of a constant curvature compact two-dimensional space, the so called horizon manifold with areal radius $\mathcal R=r$. Such a space has three different topologies, namely spherical ($k=1$), flat ($k=0$) or hyperbolic ($k=-1$). The Ricci scalar curvature reads,
\begin{eqnarray}
R=-B'' - \left(3\alpha'+ {4\over r}\right)B'- \left(2 \alpha'^2+ 2 \alpha'' + {4\over r} \alpha' +{2\over r^{2}}\right)B+{2k\over r^2} \,,
\label{Ricci}
\end{eqnarray}
where the prime denotes the derivative with respect to $r$.

Let us see which kind of SSS solutions with constant $R=R_0$ can be realized and how these solutions emerge in $F(R)$-gravity when condition (\ref{Ansatz}) is satisfied. 
Usually, $F(R)$-gravity  leads to fourth order differential equations of motion. However, if the conditions \eqref{Ansatz} hold then Eq.~(\ref{fieldequationF(R)}) in
vacuum is automatically satisfied on the solution $R=R_{0}$. At the same time,  Eq.~(\ref{Ricci}) with $R=R_0$ is a second order differential equation that can in principle be solved up to two integration constants. 

At first, we will consider the simplest case with $\alpha(r)=0$. Eq.~\eqref{Ricci} with $R=R_{0}$ can be easily solved to give
\begin{equation}
B(r)=k-\frac{c_0}{r}+\frac{c_1}{r^2}- \frac{R_0}{12}r^2\,,
  \label{ex1}
\end{equation}
where $c_0\,,c_1$ are the integration constants. 
The case with 
$0<c_1$ is equivalent to a topological
Reissner-Nordstr\"{o}m de Sitter ($0<R_0$) or Anti-de Sitter (AdS)($R_0<0$) space-time, with the only but substantial difference that $c_1$ does not correspond to the charge of
an external electric field but is purely an integration constant of the vacuum solution. 

The metric that we have found describes 
a black hole if we are able to locate an event horizon, namely a real positive value $r_{H}$ where the function $B(r)$ vanishes. In addition, we also impose that 
$B'(r_H)>0$  in order to have
positive temperature and surface gravity. For instance, if we take the spherical case $k=1$ and we neglect the contribution from the last term of $B(r)$, 
when $0<c_0$ the solution (\ref{ex1}) describes a black hole. 
Furthermore, when $0<c_1$, the Reissner-Nordstr\"{o}m black hole exhibits an internal Cauchy horizon with $B'(r_H)<0$. Recently, it has been demonstrated that such 
an horizon 
suffers from instabilities~\cite{Poisson} (see also Ref.~\cite{Maeda}), but in our case we are free to choose $c_1<0$ avoiding additional positive roots of $B(r)$. 
A similar analysis can be extended to the topological cases with
$k=0\,,-1$. Finally, the term $R_0\neq 0$ in (\ref{ex1}) is relevant at large distances and describes the asymptotic geometry of the space-time, which can be locally de Sitter 
or Anti-de Sitter.
Note that we need the AdS term with $R_0 < 0$ when $k = -1$ in order to preserve the metric signature.
In the previous chapter we have seen that the models satisfying (\ref{Ansatz}) with $R_0\neq 0$ typically describe modifications of gravity at large scales of $R\sim R_0$, 
thus we may treat the corresponding SSS metrics as local solutions in the early Universe (e.g. primordial black holes).

Note that the model $F(R)\propto R^2$ admits the solution (\ref{ex1}) for every choice of $R_0$. The reason is that the field equations 
(\ref{fieldequationF(R)}) are satisfied
for the pure dS/AdS solution and the extension to the Reissner-Nordstr\"{o}m-like case can be done thanks to the fact that the Ricci scalar, the Lagrangian $F(R)$
and its derivative $F_R(R)$
do not acquire any additional 
contribution~\cite{max,Rinaldi}.

Now, let us consider the class of the so-called Lifschitz-like solutions, namely the ones for which
\begin{equation}
\alpha (r) = \frac{1}{2} \log \left[\frac{r}{r_0} \right]^z\,.
\end{equation}
Here, $z$ is a real number and $r_0$ is a length scale. Thus, the metric (\ref{metric}) assumes the form,
\begin{equation}
ds^2= -\left( \frac{r}{r_0} \right)^z B(r)+\frac{dr^2}{B(r)}+r^2d\Omega^2_k\,.
\end{equation}
The solutions to eq.\ \eqref{Ricci} with $R=R_0$ now are
\begin{equation}\label{lifshitz total}
B(r)= K - \frac{C_{\pm}}{r^{b_{\pm}}}- \lambda r^2\,,
\end{equation}
where
\begin{align}
&K=\frac{4 k}{z^2+2 z+4} \label{K}\,, \nonumber\\
&\lambda=\frac{2 R_0}{z^2+8 z+24}\,, \nonumber\\
&b_\pm = \frac{1}{4} \left(3 z+6 \pm \frac{ \sqrt{z^4+22 z^3+48 z^2+88 z+16}}{\sqrt{z^2+2 z+4}}\right)\,,
\end{align}
and $C_{\pm}$ are the two integration constants.
Since
$(z^2+2 z+4)$ and $(z^2+8 z+24)$ are positive defined,
$K$ and $\lambda$ are positive or negative according to the signs of $k$ and $R_0$.
In order to avoid complex values of $B(r)$, we must impose that
\begin{equation}
z^4+22 z^3+48 z^2+88 z+16>0 \textbf{  }\Rightarrow \textbf{   }z \notin \left[ -2(5+2 \sqrt{6})\text{ ; }2(2\sqrt{6} -5)\right]\,.
\end{equation}
In such cases, $b_\pm$ vary as
\begin{eqnarray}
b_- \in [-\infty{ ; }\sim-13.35]\,,&\text{when}&z<-2(5+2 \sqrt{6})\,,\nonumber\\
b_- \in [\sim 0.93{ ; }+\infty]\,,&\text{when}&2(2\sqrt{6} -5)<z\,,\label{zuno}
\end{eqnarray}
and
\begin{eqnarray}
b_+ \in [-\infty{ ; }\sim-12.93]\,,&\text{when}&z<-2(5+2 \sqrt{6})\,,\nonumber\\
b_+ \in [\sim 1.35{ ; }+\infty]\,,&\text{when}&2(2\sqrt{6} -5)<z\,.\label{zdue}
\end{eqnarray}
As a check, we can see that, if we choose $z=0$, we recover the solution (\ref{ex1}) previously discussed.

Polynomial solutions of the type of (\ref{lifshitz total}) may have several real roots, namely they can provide several horizons. Typically, 
an event horizon which satisfies the condition on the positivity of $B'(r_H)$ is followed by a Cauchy horizon with $B'(r_H)<0$.
Thus, if we start with a Cauchy horizon, the second root of (\ref{lifshitz total}) will localize the event horizon of a black hole, and the external horizon will
represent the cosmological background of the solution. Furthermore, in order to prevent internal instabilities, one should avoid real and positive roots in the internal solution. 

In general, the construction of BH solutions with a single event horizon, eventually surrounded 
by a cosmological horizon, when $R_0\neq 0$, is possible for every choice of the topology. For example, if we neglect the contribution of $\lambda$, we can see that 
for the hyperbolyc topology $k=-1$ with $C_\pm<0$ a BH event horizon exists only if $b_\pm<0$ and $z$ belongs to the first interval in (\ref{zuno}) or (\ref{zdue}).
For negative and very large values of $z$ we obtain
\begin{equation}
B(r)\simeq -|K|+|C_\pm|r^{|b_{\pm}|}\,,\quad b_\pm\simeq z\,,\frac{z}{2}\,,\quad z\ll -1\,,
\end{equation}
and a single BH event horizon is located at $r_H\simeq K/C_\pm>0$.

Some of this solutions should tempt one to search for regualr BHs. For a recent review on regular BHs see e.g. \cite{Colleaux:2017ibe} and references therein.

Moreover, an important remark about the asymptotic behaviour of our BH solutions with the topological choice $k=1$ (i.e., the sphere) is in order. We observe that, if we avoid the 
contribution from the cosmological constant (in fact, we are considering the case $R_0=0$), we could not recover a flat behaviour (in particular, $-g_{00}\neq 1$). It means that 
the effects of gravity in our modified theories survive at large distances and may need some phenomenological counterterms. However, some solutions like
the Reissner-Nordstr\"{o}m metric described by (\ref{ex1}) are asymptotically flat as in GR and deserve a special interest.

We conclude this section with a general remark about the thermodynamics of our black holes. To the event horizon of a static black hole we can associate
a Killing temperature~\cite{Hawking:1976de},
\begin{equation}
T_K=\frac{\kappa_K}{2\pi}\equiv\frac{\text{e}^{\alpha(r_H)}B'(r_H)}{4\pi}\,,
\end{equation}
where  $\kappa_K$ is the Killing surface gravity.
Furthermore, the entropy of a black hole in a $F(R)$-gravity theory can be derived by
making use of the Wald formula~\cite{Wald:1993nt},
\begin{equation} 
\label{f(R) entropy}
S_W=\frac{A_k F_R(R_H)}{4G_N}\,,
\end{equation}
where $R_H$ corresponds to the Ricci scalar evaluated on the horizon at $r=r_H$ and we have reintroduced the Newton's constant $G_N$.
Here, $A_k$ is the area of the horizon,
\begin{equation}
A_k=V_k r_H^2\,,
\end{equation}
where $V_k$ depends on the topology. It turns out that, since for our class of models $F_R(R_H)=0$, the entropy of the corresponding black holes is null. It means that, 
despite to the fact that we can associate to these black holes a temperature $T_K\neq 0$, they do not emit radiation ($T dS_W=0$) so they are stable. 

In the next sections we will extend our analysis to SSS solutions for traversable wormholes and for the fitting of the observed profile of the rotation curves of galaxies.

\section{Traversable wormholes}

Traversable wormhole (WH) solutions are realized in the framework of General Relativity by making use of matter sources which violate the weak energy 
condition~\cite{flamm,ERb, Morris:1988cz,Morris:1988tu, Visser} (see also Refs.~\cite{LoboWH,LWH2,Bronnikov,Duplessis,LWH1,LWH3} and references therein). 
In $F(R)$-gravity
the role of anti-gravitational matter can be played by the modification of gravity itself and a wormhole can be obtained as a vacuum solution of the theory 
(see for instance Ref.~\cite{LWH2,Bronnikov,Duplessis,mioWH}). In this section
we will investigate a class of traversable wormhole solutions with constant or null curvature $R=R_0$ (eventually, with $R_0=0$) 
for models satisfying (\ref{Ansatz}). For our purpose we will rewrite the SSS metric (\ref{metric}) as
\begin{equation}
ds^2= -e^{2\Phi(r)} dt^2 + \frac{1}{B(r)} dr^2 + r^2 d\Omega_k^2\,,\label{metricWH}
\end{equation}
where $\Phi=\Phi(r)$ is a new redshift metric function. We should note that in spherical topology with $k=1$ and asymptotically flat geometry, 
the function $B(r)$ is usually reparameterized as $B(r)=1-b(r)/r$, $b(r)$ 
being the so-called shape function. However, in the context of a modified gravity theory, we  relax the condition on the asymptotic flat behavior of the solution and we therefore
collect the shape function and the topology inside the generic function $B(r)$.

A traversable wormhole occurs if the radial coordinate is embedded by a minimal radius, the so-called ``throat'', and no horizons appear. This means that the function $\Phi(r)$ 
must be finite and regular everywhere along the throat. Specifically,
the following
conditions must be satisfied (see Refs.~\cite{Morris:1988cz,Visser,LoboWH,Visser:1997yn,Hochberg:1997wp,Visser:1989kh}):
\begin{itemize}
\item $\Phi(r),B(r)$ are smooth functions for all $r \geq \tilde r$;
\item $\Phi'_+(\tilde r)=\Phi'_-(\tilde r)$;
\item $B(\tilde r) = 0$ and   $B(r)>0\quad \forall \quad r>\tilde r$;
\item  $B'_+(\tilde r) = B'_-(\tilde r)>0$.
\end{itemize}
Here, $r=\tilde r$ localizes the radius of the throat of the wormhole. The proper distance is defined by
\begin{equation}
l(r)=\pm\int^{r}_{\tilde r}\frac{1}{\sqrt{B(r')}}d r'\,, 
\end{equation}
and the areal radius of the horizon reaches a minimum when $l=l(\tilde r)$. The positive and negative values of $l$ correspond to 
the lower and upper universes connected by the wormhole.
Thus, the coordinate time required by a traveler to cross the wormhole
between $l=-l_1<0$ and $l=+l_2>0$ is given by,
\begin{equation}
\Delta  t=\int ^{l_2}_{l_1} \frac{1}{v e^{\Phi(l)}} d l\,, \label{timetravel}
\end{equation}
where $v=d l/[\text{e}^{\Phi(l)}dt]$ is the radial velocity of the traveler as he/she passes radius $r$  ~\cite{Morris:1988cz}.
It is clear that the black hole solutions found in the previous section cannot describe traversable
wormholes since the redshift function $\Phi(r) $ is singular on the throat (horizon). We should also note that, when $\Phi'(r)<0$, a 
repulsive tidal force arises, which may make impossible to travel across the throat. 

The Ricci scalar for the metric (\ref{metricWH}) reads,
\begin{align} 
\label{WHRicci}
R=-\frac{B'}{r} \left(r \Phi '+2\right)-\frac{2 B}{r^2} \left(r^2 \Phi ''+r^2 \Phi '^2+2 r \Phi '+1\right)+\frac{2 k}{r^2}\,.
\end{align}
In what follows, we will check for some solutions satisfying the equation $R=R_0$ by choosing specific functional forms of $\Phi(r)$. 

Let us start with the simplest case $\Phi (r)= \text{const}$, namely we consider a vanishing tidal force. The proper time measured by a static observer at $r$
corresponds to the coordinate time $t$.
For simplicity, and without loss of generality, we set $\Phi (r)=0$. Then, from (\ref{WHRicci}) with $R=R_0$ we derive
\begin{equation}
\label{z=0}
  B(r)= k- \frac{c_0}{r}-  \frac{R_0\,r^2}{6}\,,
\end{equation}
where $c_0$ is a constant. Thus the $g_{rr}(r)$ component of the metric is the same as in the Schwarzschild-dS/AdS solution. This solution clearly describes a traversable wormhole, being the integral in (\ref{timetravel}) convergent. For example, if $R_0>0$, 
we have solutions with $k=0$, $c_0>0$ and $\tilde r= \sqrt[3]{\frac{-6c_0}{R_0}}$.\\

Let us consider now the class of models
\begin{equation}
\Phi(r)= \frac{1}{2} \log \left[ \left( \frac{r}{r_0}\right)^z\, \right]\,,
\end{equation}
where $z$ is a real number and $r_0$ some length scale. The metric (\ref{metricWH}) becomes
\begin{equation}
ds^2= -\left( \frac{r}{r_0}\right)^z  dt^2 + \frac{1}{B(r)} dr^2 + r^2 d\Omega_k^2\,,\label{WH2}
\end{equation}
and the equation $R=R_0$ yields
\begin{equation} 
B(r)=K- \frac{c_0} {r^{b}}- r^2\lambda\,,\label{WH22}
\end{equation}
where $c_0$ is a constant and
\begin{align}
&K=\frac{4 k}{z^2+2 z+4} \,,\nonumber\\
&b=\frac{z^2 + 2z+4}{z +4} \,,\nonumber\\
&\lambda=\frac{2 R_0}{z ^2+4 z +12}\,.
\end{align}
Note that (\ref{WH22}) reduces to (\ref{z=0}) when $z=0$. With the metric 
(\ref{WH2})--(\ref{WH22}) we can describe traversable wormholes for every choice of the topology $k$, since the metric 
function $\Phi(r)$ is well defined for $r>0$ (and therefore for $ r> \tilde r \,, \tilde r>0$). 
When $b=2$, namely $z=\pm 2$, and $\lambda=0$ we get a Morris-Thorne-like traversable wormhole~\cite{Morris:1988cz}. In this case we have
\begin{equation}
l=\pm \sqrt{r^2-\tilde{r}^2}\,,
\end{equation}
and the metric reads
\begin{equation}
ds^2=  - \frac{(l^2+\tilde{r}^2)^{\pm 1}}{r_0^{\pm 2}}  dt^2 +  dl^2 + (l^2+\tilde{r}^2) d\Omega_k^2\,,
\end{equation}
where the plus/minus signs correspond to the choices $z=+2$ and $z=-2$, respectively. 
For instance, for the spherical topology with $k=1$, the throat of the wormhole is located at $\tilde r= \sqrt{c_0/K}$. Note that
$K=1$ when $z=-2$ and $K=1/3$ when $z=2$.
For the generalization of the Morris-Thorne wormhole solution in the presence
of a cosmological constant (here, with $R_0\neq 0$ such that $\lambda\neq 0$), see Ref.~\cite{LoboWH}.

We conclude this section by discussing a 2-parameter family of wormholes. 
Let us consider the metric 
\begin{equation}
\label{WHlast}
 ds^2=-{A}(r)d t^2+\frac{dr^2}{{B}(r)}+r^2d\Omega^2_k\,,
\end{equation} 
where
\begin{eqnarray} 
A(r)&=& k-c_0/r\,,\nonumber\\
B(r) &=&\frac{A(r)}{k+ 3 A(r) }  \left(4 k -\frac{c_1}{r}  - \frac{2}{3} r^2 R_0 \right)\,,
\end{eqnarray}
and $c_{0}$, $c_{1}$, $R_{0}$ are constant. The parameter $k$ is the usual topological index for the horizon space. One can show that the Ricci scalar is constant and $R=R_0$. If $c_0=0$, after a rescaling of the time coordinate, we recover the solution (\ref{z=0}).
The topological case $k=-1$ is meaningless, since the metric will not preserve the signature for large values of $r$. On the contrary, for $k=0$ and $c_0<0$, a traversable wormhole solution may be obtained when $c_1>0$ and $R_0<0$. Finally, for $k=1$ a traversable wormhole solution can be realized only if $B(\tilde r)=0$ when $A(\tilde r)>0$. For example, if $c_0>0$ (such that at large distance the solution exhibits a Newtonian potential), we must require that $A(r)=0$ only if $r<\tilde r$. In the case $R_0=0$ we could have traversable solutions for $0<4c_0<c_1$.

\section{Rotation curves of galaxies}

SSS metrics have useful applications in the analysis of the rotation curves of galaxies. 
In the context of Einstein's gravity, in first approximation, galaxies are characterized by two ingredients: an internal disk composed of visible baryonic matter, 
and a spherical halo of dark matter. The latter is necessary in order to explain the observed flattening of the rotation curve at large distance from the galactic center.  A description of the gravitational active content of galaxies can be found in Refs.~\cite{Salucci:2007et, gal in univ}. From the phenomenological point of view, they lead to a Newtonian potential which brings to~\cite{Riegert:1984zz, Mannheim:1988dj, Mannheim:2005bfa, Mannheim:2010ti, Mannheim:2010xw, OBrien:2011vks, Capozziello:2006ph, Salucci:2014oka, Salucci:2013rmp, Hashim:2014mka},
\begin{equation}
g_{00}(r) = -\left(1+ 2\phi_{\rm tot}(r)\right) = -\left(1+2\phi_{\rm BM}(r)+2\phi_{\rm DM}(r) \right) = -\left( 1-\frac{c_0}{r}+ c_1 r \right)\,,
\end{equation}
where $c_0\,,c_1$ are positive dimensional constants
and
$\phi_{\rm tot}(r)$ is the total potential given by the baryonic matter potential $\phi_{\rm BM}(r)$ and the dark matter potential $\phi_{\rm DM}(r)$.
Note that the baryonic matter potential corresponds to the classical Newtonian potential, while the dark matter potential is linear with respect to the radial coordinate 
and reproduces the observed flattening of the rotation curves.

If we neglect the behaviour for small $r$, we can derive the interpolating form~\cite{Roberts:2002ei},
\begin{equation}
-g_{00}(r) =1+ c_2 \ln(r)\,,\label{g1}
\end{equation} 
$c_2$ being a positive constant, while at large distances we simply have,
\begin{equation}
-g_{00}(r)=1+c_1 r\,. \label{g2}
\end{equation}
In modified gravity,  dark matter does not correspond to a fluid of particles but rather an effect of the geometry, induced by a deviation from standard GR. Thus, we wish to explore the possibility that the models satisfying the Ansatz (\ref{Ansatz})  reproduce the correct profile of the rotation curves. We consider again the metric with spherical topology ($k=1$) 
\begin{equation}
ds^2=-A(r)dt^2+\frac{dr^2}{B(r)}+r^2d\Omega_1^2\,, 
\end{equation}
such that the Ricci scalar is constant. In particular, we study the case 
\begin{equation}\label{bare}
A(r)=1+c_2 \ln(r)\,,
\end{equation}
which accounts for 
the contribution of both visible and dark matter as in
(\ref{g1}), and  the case
\begin{equation} \label{DM only}
A(r)=1+c_1 r\,,
\end{equation}
which describes the asymptotic behavior of dark matter only as in (\ref{g2}).

In the first case, one finds that solutions with constant scalar Ricci curvature $R=R_0$ lead to the following implicit form of $B(r)$,
\begin{equation}
B(r)=\frac{\left(c_3+\beta (r)\right)\text{e}^{-\frac{9}{4} \tanh ^{-1}\left(1+\frac{8}{c_2 }+
8 \ln(r)\right)}}{r \sqrt[8]{(4 +c_2 )+c_2  (c_2 +8 ) \ln(r)+4 c_2 ^2 \ln^{2}(r)}}\,,
\end{equation}
where $c_3$ is an integration constant and
\begin{equation}
\beta(r) = \int_1^r \frac{2 \left(2 -R_0 \tilde{r}^2\right) (1+c_2  \ln
   (\tilde{r}))^2\text{e}^{-\frac{9}{4} \tanh ^{-1}\left(1+\frac{8}{c_2 }+
8 \ln (\tilde r)\right)}}{\left(c_2  (c_2 +8 ) \ln (\tilde{r})+4 c_2 ^2 \ln
   ^2(\tilde{r})+ (4 +c_2 )\right)^{7/8}} \, d\tilde{r}\,.
\end{equation} 
In the second case, we get
\begin{equation} \label{dm}
B(r)=\frac{1+c_1  r}{357 r} \left(\frac{357 c_4}{(4 +5 c_1  r)^{7/5}}-\frac{2 \left(8  R_0-2 c_1   (51 c_1 +7 r R_0)+21 c_1 ^2 r^2   R_0\right)}{c_1 ^3}\right)\,,
\end{equation}
where $c_4$ is an integration constant.
This solution gets extremely simplified if we consider $R_0=0$ and $c_4=0$,
\begin{align}
B(r)= \frac{4(1+c_1  r)}{7c_1 r}=  \frac{4}{7}\frac{ A(r)}{c_1 r}\,.
\end{align}
As we have seen in the second section, $F(R)$-models satisfying (\ref{Ansatz}) with $R_0=0$ are largely used to reproduce the accelerated expansion of our Universe.
Thus,
it is remarkable to observe that this kind of models for the dark energy may also take into account the dark matter at the galactic scale providing some simple solutions. 

In any case, speculations about alternative dark matter models, as the ones described in Refs.~\cite{max2, recflat, Shojai:2013uva} or the one  shown above,
need a careful investigation after the data from the neutron star merging  GW170817 \cite{TheLIGOScientific:2017qsa} (see the analysis in Ref.~\cite{Boran:2017rdn}). 
In the specific, the MOND theories are critically questioned, due to the fact that
the velocity of the gravity waves is near to the velocity of light. However, the particle matter theories (see for example Ref.~\cite{Lu:2006ah}) are 
still unaffected by the GW170817 (for example, the models in Refs.~\cite{max2, recflat, Shojai:2013uva} may actually provide a viable alternative to dark matter).

\section{Conclusions}

\noindent In this paper, we have studied spherically symmetric solutions in a special class of vacuum $F(R)$ gravity. These solutions are characterized by a constant (eventually vanishing) Ricci scalar and are suitable to describe compact objects such as black holes and wormholes. In addition, they might describe also the anomalous rotation curve typical of spiral galaxies. 

The main lesson is that the space of solutions is  very large and worth studying since it provides a rich phenomenology. In the case of black holes, the metric is often different from the standard Schwarzschild one thus it is possible to compare our results with Solar System tests and constrain the parameter space. Thus it would be very interesting  to carry on these investigations by applying the PPN formalism necessary to this kind of experimental verifications. Similarly, the study of gravitational waves coming from black hole mergers might provide further constrains on the parameters. 

As a byproduct of these results, we have found that these models can also mimic the influence of the dark matter halo that is supposed to surround spiral galaxies. Also in this case,  there is a potential for fixing the  parameters of the theory by analyzing the rotation curves of a large number of galaxy.

Concerning the wormholes, we found few configurations that lead to traversable ones. The conceptual advantage of these solutions is the fact that there is no need for unnatural matter sources that violate GR energy conditions. Nevertheless, as for the black hole solutions, nothing can be said, at this level, about the precise functional form of $F(R)$ except that it must satisfy the constraints $F(R_0)=0,\,\,\, F_R(R_0)=0$. The latter conditions might be the result of some  fundamental symmetry of vacuum $F(R)$ gravity that select only the solutions with constant or vanishing Ricci scalar. What this symmetry might be will be the subject of future studies.


\end{document}